# Secure internal communication of a TrustZone-enabled heterogeneous SoC lightweight encryption

El Mehdi Benhani[1], Cuauhtemoc Mancillas Lopez[2], Lilian Bossuet[1], *Senior Member, IEEE*

1 Hubert Curien Laboratory University of Lyon, Saint-Etienne, France
2 Computer Science Department, CINVESTAV-IPN, Mexico
Emails: elmehdi.benhani@univ-st-etienne.fr,
lilian.bossuet@univ-st-etienne.fr
cuauhtemoc.mancillas@cinvestav.mx

*Abstract—* Security in TrustZone-enabled heterogeneous system-on-chip (SoC) is gaining increasing attention for several years. Mainly because this type of SoC can be found in more and more applications in servers or in the cloud. The inside-SoC communication layer is one of the main element of heterogeneous SoC; indeed all the data goes through it. Monitoring and controlling inside-SoC communications enables to fend off attacks before system corruption. In this article, we study the feasibility of encrypted data exchange between the secure software executed in a trusted execution environment (TEE) and the secure logic part of an heterogeneous SoC. Experiment are done with a Xilinx Zynq-7010 SoC and two lightweight stream ciphers. We show that using lightweight stream ciphers is an efficient solution without excessive overheads.

## I. INTRODUCTION

Modern heterogeneous SoC (such as Xilinx Zynq or Intel Cyclone V SoC FPGA) has two parts; a processing system part includes hard components and ARM cores, and a programmable logic part for custom designs. Both parts include the ARM TrustZone technology which is a hardware security extension that helps to partition both hardware and software resources into two worlds, one secure where high-value code and data can be protected and one non-secure. The communications in this heterogeneous SoC is ensured using the system bus like the Advanced eXtensible Interface (AXI) bus. From a SoC vendor to another, the name of the two parts and the communication bus can change, but the TrustZone technology implementation is the same.

The security of the communication between the two parts of a heterogeneous SoC is crucial, particularly when the system uses sensible data. Previous works [1, 2] showed that a small hardware Trojan inserted in the custom design implemented in the programmable logic can compromise not only the AXI bus communication but the entire system. The hardware Trojan give access to a non-secure software to a secure hardware IPs. Moreover, a malicious FIFO inserted in the AXI Interconnect (the IP in charge of managing the bus communication) stole valuable data and shared it with non-secure IPs.

In the literature, several works tackled the security of bus communication. The work in [1, 3, 4, 5, and 6] proposed to check the security rules directly on the bus in order to not degrade the system performances by useless external memory access. Supposing that the entire SoC is trusted, Brunel et al. [7] proposed Secbus to protect the SoC from the attacks that targeted the external memory. In this article, we study for the first time the feasibility of using lightweight encryption algorithms in order to protect inside-SoC sensitive communication. Unlike [7], we suppose that the heterogeneous SoC is not completely trusted and we include threats like the one presented in [1, 2].

The paper is organized as follows: we present the encryption scheme and the two lightweight stream ciphers (Trivium and Grain-128a) in Section 2, the software and hardware performance results with a presentation of the Xilinx Zynq-7010 experimental setup in Section 3. Section 4 concludes the paper.

## II. BUS ENCRYPTION:

In this section, we present the implementation of the encryption scheme and the two lightweight stream ciphers, Trivium and Grain-128a.

### A- SoC architecture with secure partition

The first step for implementing the encryption scheme is to make an inside-SoC secure partition of the SoC processing system and of the SoC programmable logic both as presented in figure 1. The creation of a secure partition of the programmable logic should follow the methodology presented in [1]. The methodology makes use of TrustZone configuration registers and Xilinx Isolation Design Flow (IDF). The second step is to include cryptographic elements in the secure world in the SoC processing system and in the SoC programmable logic both. To do it, a crypto trusted application (TA) is load in the trusted execution environment (TEE) of the SoC processing system, and a crypto IP is configured in the secure hardware area of the programmable logic.

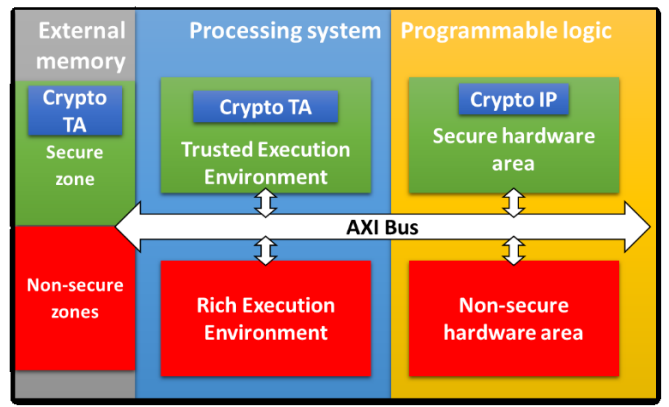

Figure 1: Secure partition of a TrusZone-enabled heterogeneous SoC

The crypto IP has four blocks as presented in figure 2. An AXI master interface used to communicate with the targeted IP in the secure hardware area using decrypted data. An AXI slave interface used to exchange encrypted data with the crypto TA. A controller used for pre-processing, post-processing, and scheduling the encryption and the decryption operations. A lightweight stream cipher to encrypt and decrypt sensitive data.

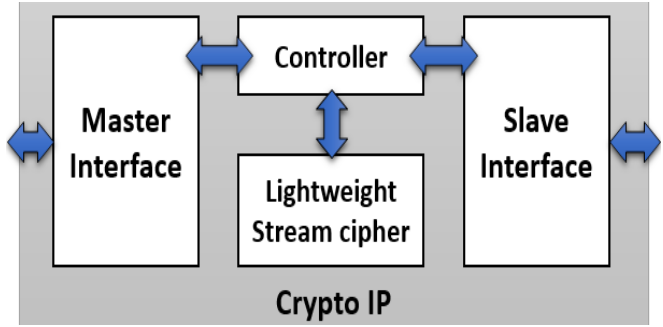

Figure 2: Crypto IP architecture

The crypto TA executes a lightweight stream cipher for sensitive data encryption and decryption. The crypto TA is included in the processing system TEE presented in figure 3. The crypto TA is used any time a trusted application want to exchange data with a secure Hardware IP.

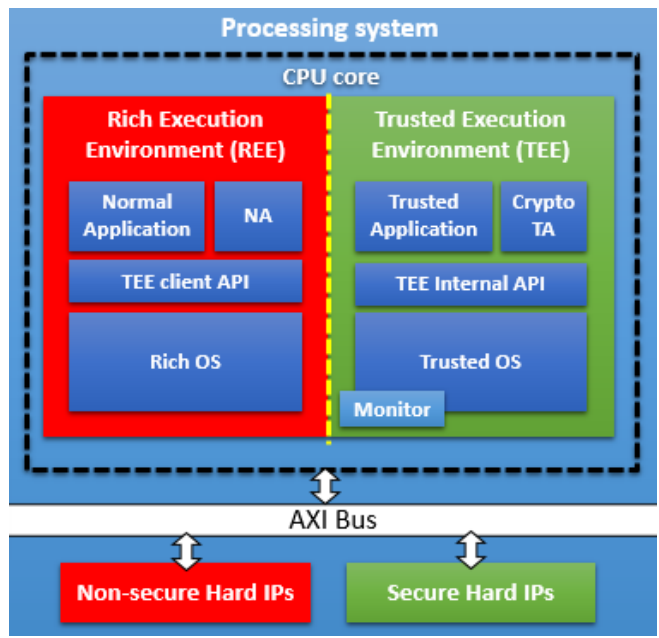

Figure 3: Processing system TEE

### B- Stream ciphers, Grain-128a, Trivium:

Stream ciphers are widely known cryptographic primitives that allow to perform bulk encryption, they receive as input a key K and an initialization vector (IV) and outputs a string of pseudorandom bits (stream) which is added using a logical Xor operation with the information to be encrypted. The decryption process is analogous to encryption, the same K and the same IV are necessary to generate the same stream cipher in order to Xor it with decrypted data to recover the original message. We used two lightweight stream ciphers from the hardware portfolio of eSTREAM project [8]: Grain128a [9] and Trivium [10]. Below, we provide a detailed description of both stream ciphers.

**Grain-128a:**

Grain-128a is a synchronous lightweight stream cipher an improvement of Grain-128, it added security enhancements and optional message authentication using the Encrypt & MAC approach. It has two main parts: Pre-output function and a MAC. The pre-output function has an internal state size of 256 bits, consisting of two registers of size 128 bit: NLFSR (b) and LFSR (s). The MAC supports variable tag lengths w such that $0 < w \leq 32$. The cipher uses a 128-bit key and an Initialization Vector (IV) of 96 bits for the initialization process. Grain-128a is part of the Grain family of stream ciphers; its throughput can be increased at the expense of additional hardware.

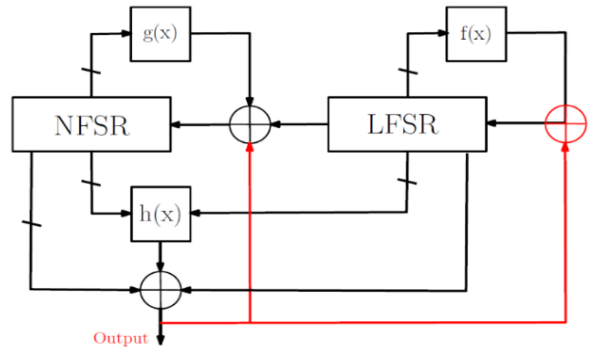

Figure 4: Pre-output of Grain-128a

**Trivium:**

Trivium is a synchronous stream cipher designed to provide a flexible trade-off between speed and gate count in hardware. It uses 80-bit key and an 80-bit IV. It has an internal state of 288 bits. The internal state consists of three shift registers of different lengths as presented in Figure 5. At each round, a bit is shifted into each of the three shift registers using a non-linear feedback function with input from that and one other register, in the same time one bit of output is produced.

To initialize the cipher, the key and initialization vector are written into two of the shift registers, with the remaining bits starting in a fixed pattern, the cipher state is then updated 1152 times so that every bit of the internal state depends on every bit of the key and of the IV in a complex nonlinear way.

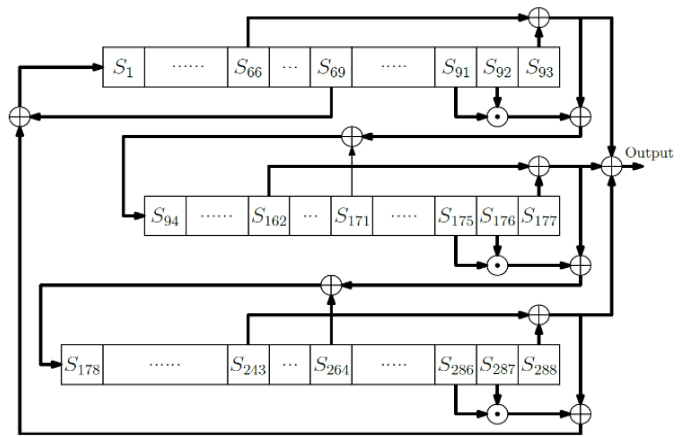

Figure 5: Trivium

### III. SOFTWARE AND HARDWARE PERFORMANCE RESULT:

In this section, we present the target platform Zynq-7010, followed by the software and hardware experimental results.

### A- The Xilinx Zynq-7010 setup:

The Xilinx Zynq-7010 is a TrustZone-enabled heterogeneous SoC. During the experiments, the dual ARM cortex-A9 in the processing system is running at 600MHz, and the custom design included in the programmable design at 200MHz, the external memory is partitioned into 128MB for the secure world and 384MB for the non-secure world. We implemented the encryption scheme presented above using a custom TEE, and custom design with secure and non-secure IPs.

Interested readers can follow the online tutorial [11] on designing a TrustZone-enable system with the Xilinx Vivado CAD tool. And check [1] for more detail about securing a TrustZone-enable heterogeneous SoC.

### B- Software performance of the crypto TA:

We took the source code of Trivium and Grain-128a from the reference implementation submitted to eSTREAM [8] and then we have modified them to run on the Xilinx Zynq-7010. Table 1 shows the size in bits of the different parameters used for each stream cipher. We use a counter synchronized with the crypto IP as generator of IVs.

Table 1: Lightweight stream cipher RAM size

| Stream cipher | Key size (bits) | Max IV size (bits) | Internal state size (bits) |
|---|---|---|---|
| Grain-128a | 128 | 96 | 256 |
| Trivium | 80 | 80 | 288 |

We modified the reference implementations (originally bit-oriented) to raise the output rate and speed up the data encryption and decryption. We implemented 8-bit, 16-bit and 32-bit output rates for both algorithms (the Zynq-7010 data path is limited to 32-bit). Figure 6 shows the execution time of each implementation of both algorithm processing 1MB in the Zynq-7010.

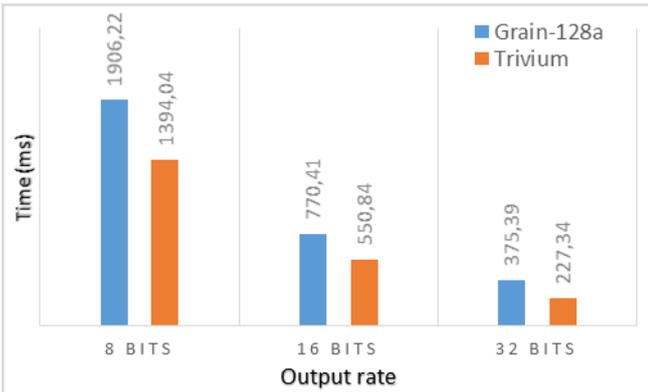

Figure 6: Execution time of different implementation of Grain-128a and Trivium to encrypt 1MB of data.

In Figure 6 it is clear that Trivium is faster than Grain-128a; this is because the feedback and output functions used by Trivium are less complex than ones use by Grain-128a. More complex functions mean more clock cycles for their computation on the processors.

### C- Hardware performance of the crypto IP:

In the programmable logic, we implemented the different elements of the crypto IP without the stream cipher block using approximately 105 LUTs. The lightweight stream cipher blocks were designed based on [9, 10]. Like software implementations, we implemented Trivium and Grain-128a with different output rate (8-bit/cycles, 16-bit/cycles, and 32-bit/cycles). Figure 7 shows the initialization time in clock cycles taken by each algorithm at the beginning of each encryption or decryption. Grain-128a is clocked 256 times and Trivium 1152 times without produce any output bit. The initialization time decrease linearly with the output rate.

Figure 8 and 9 show the number of Flip-Flop and the number of LUTs required by each implementation of Trivium and Grain-128a. The complexity of functions used by Grain-128a implies the necessity of more LUTs than for Trivium which optimizes the number of LUTs. It is directly related to the number of variables in each function.

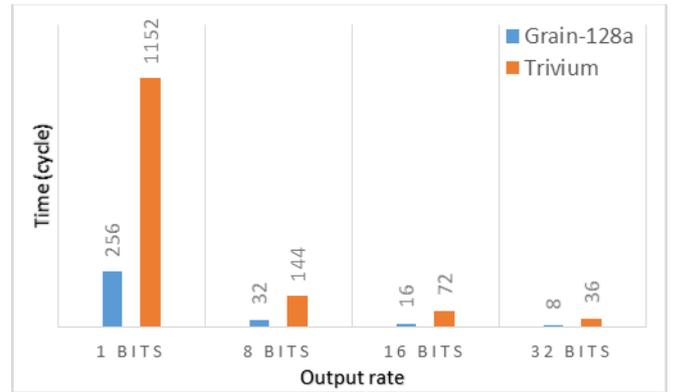

Figure 7: Initialization time of Trivium and Grain-128a

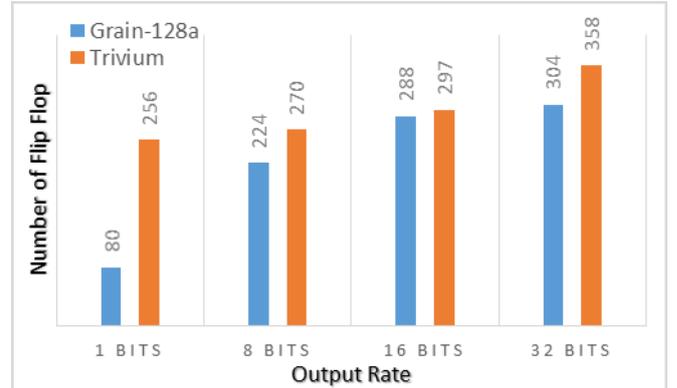

Figure 8: Number of flip flop occupied by Trivium and Grain-128a

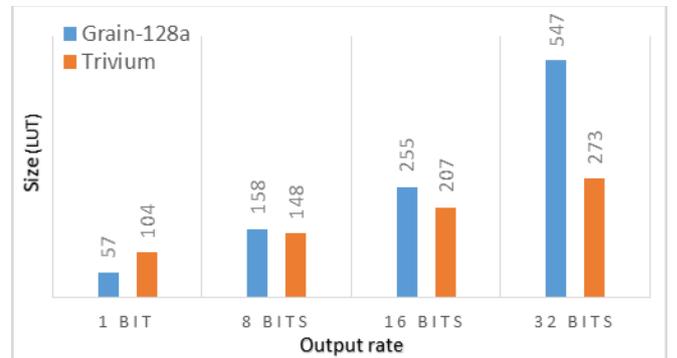

Figure 9: Size in LUTs of Trivium and Grain-128a

### D- Introduced overhead:

The timing overhead introduced by the encryption of the AXI bus in the programmable logic is caused by the initialization process of the stream ciphers, using the 32-bit data-path size it represents 8 and 36 clock cycles for Grain-128a and Trivium respectively. In the processing system, we have measured the execution time of both stream ciphers implementations which is equal to 15.3us. Such timing overhead are not excessive because the encryption/decryption are used only to protect sensitive data processed by the secure world. Most of the case, the applications are run in the normal world with not any encryption/decryption

The memory space/area overheads are not excessive too. The memory space range required by the crypto TA is 76kB to 77KB. That is pretty small compared to the available memory (128MB for the secure memory zone in our case). The area range in number of slices required by the crypto IP is 162 to 697. That equivalent to 0.93% to 3.47% of the Xilinx Zynq-7010. When the lightweight cipher data-path is limited (from 1 to 8 bits) the area overhead is very limited and allows the designer to protect sensitive data efficiently.

## IV. ACKNOWLEDGEMENT:

The work presented in this paper was realized in the frame of the Archi-Sec project supported by the French "Agence Nationale de la Recherche"

## V. CONCLUSION:

Inside-SoC communication security is mandatory to protect sensitive data in heterogeneous SoC even when security mechanism like ARM TrustZone is used. In this paper we show that using lightweight stream cipher is an efficient solution to secure the communication channel between a TEE running in the processing system of the SoC and the secure hardware area in the programmable logic. Such work conduce to think that FPGA companies should place crypto-resources in their heterogeneous SoC to guarantee security and performance.